\RequirePackage{fix-cm}
\documentclass[smallextended]{svjour3}       % onecolumn (second format)
\smartqed  % flush right qed marks, e.g. at end of proof
\usepackage[authoryear]{natbib}
\usepackage{graphicx}
\usepackage{hyperref}
\usepackage{xcolor}
\definecolor{darkgreen}{rgb}{0.0, 0.5, 0.0}

\newcommand{\rqone}{How effective are fine-tuned LLMs in SATD identification and classification?}
\newcommand{\rqtwo}{Does our proposed ICL with a larger model outperform smaller models that have been fine-tuned in identifying and classifying SATD?}
\newcommand{\rqthree}{What is the impact of adding the classification layer in fine-tuning LLMs?}
\newcommand{\rqfour}{What is the impact of additional contextual features on LLM-based SATD classification?}

\begin{document}

\title{An Empirical Study on the Effectiveness of Large Language Models for SATD Identification and Classification}
% \subtitle{Do you have a subtitle?\\ If so, write it here}

\titlerunning{LLM for SATD Identification and Classification}        % if too long for running head

\author{Mohammad Sadegh Sheikhaei         \and
        Yuan Tian   \and
        Shaowei Wang \and
        Bowen Xu
}


\institute{Mohammad Sadegh Sheikhaei and Yuan Tian \at
              School of Computing, Queen's University, Kingston, ON, Canada \\
             \email{sadegh.sheikhaei@gmail.com \quad y.tian@queensu.ca}
           \and
           Shaowei Wang \at
              Department of Computer Science, University of Manitoba, Winnipeg, MB, Canada \\
              \email{Shaowei.Wang@umanitoba.ca}
           \and
           Bowen Xu \at
              Department of Computer Science, North Carolina State University, Raleigh, NC, US \\
              \email{bxu22@ncsu.edu}
}

\date{Received: date / Accepted: date}

\maketitle

\begin{abstract}
Self-Admitted Technical Debt (SATD), a concept highlighting sub-optimal choices in software development documented in code comments or other project resources, poses challenges in the maintainability and evolution of software systems. Large language models (LLMs) have demonstrated significant effectiveness across a broad range of software tasks, especially in software text generation tasks. Nonetheless, their effectiveness in tasks related to SATD is still under-researched.
In this paper, we investigate the efficacy of LLMs in both identification and classification of SATD. For both tasks, we investigate the performance gain from using more recent LLMs, specifically the Flan-T5 family, across different common usage settings. %These settings include in-context learning (ICL), fine-tuning, and the modification of the last layer with a classification layer during fine-tuning. Additionally, we study the impact of providing contextual information on the SATD classification.

%The larger model, Flan-T5-XL with 2.85B parameters, achieved marginally better results, with a 0.4\% to 2.4\% higher F1 score compared to its smaller counterparts with 77M to 783M parameters.
%Replacing the text generation layer with a classification layer in Flan-T5 models does not benefit SATD identification but enhances SATD classification performance, particularly in smaller model versions, likely due to limited training data for SATD classification.
%Conversely, the ICL approach and smaller fine-tuned LLMs experience a decline in performance when presented with complex contextual information.
Our results demonstrate that for SATD identification, all fine-tuned LLMs outperform the best existing non-LLM baseline, i.e., the CNN model, with a 4.4\% to 7.2\% improvement in F1 score. In the SATD classification task, while our largest fine-tuned model, Flan-T5-XL, still led in performance, the CNN model exhibited competitive results, even surpassing four of six LLMs. 
We also found that the largest Flan-T5 model, i.e., Flan-T5-XXL, when used with a zero-shot in-context learning (ICL) approach for SATD identification, provides competitive results with traditional approaches but performs 6.4\% to 9.2\% worse than fine-tuned LLMs. For SATD classification, few-shot ICL approach, incorporating examples and category descriptions in prompts, outperforms the zero-shot approach and even surpasses the fine-tuned smaller Flan-T5 models. Moreover, our experiments demonstrate that incorporating contextual information, such as surrounding code, into the SATD classification task enables larger fine-tuned LLMs to improve their performance. Our study highlights the capabilities and limitations of LLMs for SATD tasks and the role of contextual information in achieving higher performance with larger LLMs, setting a foundation for future efforts to enhance these models for more effective technical debt management.

\keywords{Self-admitted technical debt (SATD) \and SATD identification \and SATD classification \and Large language models \and Fine tuning \and In-context learning}
\end{abstract}

\section{Introduction}\label{sec:introduction}
Technical Debt (TD) refers to the deliberate adoption of less-than-ideal solutions in software design or coding, typically to meet urgent deadlines or address immediate resource constraints~\citep{Cunningham-1992}. This metaphor portrays the trade-off between short-term expediency and long-term software quality, likening it to incurring financial debt that accrues interest in the form of increased maintenance effort over time~\citep{Buschmann-2011}. In literature, large-scale analysis on TD is often facilitated through the study of one specific type of TD, i.e., Self-Admitted Technical Debt (SATD)~\citep{Potdar-2014}. SATDs are sub-optimal choices that developers consciously make and primarily document in code comments~\citep{Maldonado-2015,Guo-2021,OBrien-2022}, though developers may also occasionally record them in other software artifacts such as issue reports~\citep{Li-2022}.

Addressing SATD is crucial, as its accumulation can significantly impair the long-term maintainability and evolution of software systems~\cite{Li-2023-estimating-effort}. A study by Wehaibi et al.~\citep{wehaibi2016examining} indicates that source code files with SATD experience a higher frequency of bug-fixing changes than those without SATD. To tackle this challenge effectively, software development teams must manage SATD by first pinpointing the locations of existing SATDs within their software (\textit{SATD Identification})~\citep{Sheikhaei-2023}. Subsequently, these SATDs could be categorized into various types based on their specific impacts on software maintenance, such as design, requirement, defect, documentation, and test debt (\textit{SATD Classification}). This categorization aids in prioritizing efforts and efficiently allocating developers to address and resolve these SATDs~\citep{Maldonado-2017}. However, manually performing the above two tasks, i.e., SATD identification and SATD classification, is challenging and time-consuming. For SATD identification, in large projects, only approximately 0.5 to 4\% of a project's code comments are SATD~\citep{Guo-2021}. While certain keywords like TODO and FIXME strongly suggest a code comment as SATD~\citep{[Rantala-2020,Guo-2021}, developers often mention technical debt without using these specific keywords. Consequently, only about 20 to 90\% of SATD in a project can be identified using such keywords~\citep{Yu-2022}. For SATD classification, there exist no specific keyword sets that can be universally applied to effectively detect the various types of SATDs. This limitation underlines the need to develop sophisticated automated approaches to identify and categorize SATDs.

There have been numerous studies on automated SATD identification and classification~\citep{Maldonado-2015,Maldonado-2017,Ren-2019,Guo-2021,Xiao-2021,Cassee-2022,Yu-2022,OBrien-2022,Sridharan-2023}. These approaches are either rule-based or machine learning-based. While these models show promise, their performance on these two classification tasks could be further enhanced. Recently, machine learning has seen the emergence of Large Language Models (LLMs). LLMs have exhibited strong capabilities in text generation tasks, such as text summarization and question answering~\citep{Naveed-2023}. In the software engineering (SE) domain, researchers also found that LLMs can achieve state-of-the-art (SOTA) performance on generative tasks such as code generation~\citep{wei2023magicoder}, code summarization~\citep{yuan2023evaluating}, and program repair~\citep{jin2023inferfix}. However, to the best of our knowledge, no prior studies have investigated how to better use these complex models specifically for SATD identification and classification. Furthermore, LLMs have shown impressive capabilities in in-context learning (ICL)~\citep{Brown-2020}. With the ICL ability, an LLM can adapt to new tasks or understand new information by leveraging examples provided directly in its input (prompt) without fine-tuning - a costly and time-consuming process that adapts pre-trained models to specific tasks due to updating model parameters. This raises an intriguing question: can ICL with LLMs achieve competitive performance in SATD identification and classification compared to fine-tuned LLMs?

In this paper, we aim to further tap into the potential of LLMs in two important SATD-related tasks, i.e., SATD identification and classification. We explore the impact of various usage settings of LLMs in these tasks, encompassing fine-tuning versus ICL, model size, prompt engineering, and adaptation in model architecture. Additionally, we explore the potential of integrating contextual features beyond code comment into SATD classification. Our findings can guide future research and advance the use of LLMs in managing SATD and other classification tasks within the SE field.

To achieve our goal, we conduct an empirical study using two well-known datasets: the Maldonado-62k dataset~\citep{Maldonado-2017} (we use the revised version by~\citep{Yu-2022}), and the OBrien dataset~\citep{OBrien-2022}. The Maldonado-62k dataset comprises 62,275 code comments from ten popular open-source software projects across various application domains, of which 7.2\% (4,497 comments) are identified as SATD. The OBrien dataset includes 856 SATDs randomly sampled from 68,820 SATDs extracted from 2,641 popular machine-learning repositories on GitHub. These SATDs are categorized into six general types: requirement, code, test, defect, design, and documentation. Our empirical study answers the following four research questions (RQs): 

\vspace{0.1cm}
\noindent \textbf{RQ1 \textit{\rqone}} We evaluated the performance of six large language models: BERT-base~\citep{Devlin-2019}, CodeBERT~\citep{Feng-2020}, and four variants of Flan-T5 (small, base, large, and XL)~\citep{Chung-2022}, comparing them against existing baselines for the SATD identification and classification. Our analysis revealed that for SATD identification, the selected LLMs outperformed the best existing baseline, showing a 4.4\% to 7.2\% improvement in F1 score. The larger model, Flan-T5-XL, achieved marginally better results, with a 0.4\% to 2.4\% higher F1 score compared to its smaller counterparts. In the SATD classification task, while the fine-tuned Flan-T5-XL still led in performance, the CNN model exhibited competitive results, even surpassing four of six LLMs.

\vspace{0.1cm}
\noindent \textbf{RQ2 \textit{\rqtwo}} We propose an ICL approach for SATD identification and classification using the largest Flan-T5 model, i.e., Flan-T5-XXL~\citep{Chung-2022}. We found that a zero-shot approach with the Flan-T5-XXL model provides competitive results for SATD identification with traditional approaches but performs 6.4\% to 9.2\% worse than fine-tuned LLMs. For SATD classification, few-shot incorporating examples and category descriptions in prompts outperforms the zero-shot approach and even surpasses the results of fine-tuning smaller Flan-T5 models, i.e., the small and base versions. 

\noindent \textbf{RQ3 \textit{\rqthree}} Different LLM architectures naturally lend themselves to distinct fine-tuning strategies. We found that in the absence of sufficient training data, changing the architecture of the Flan-T5 models by substituting its original text generation layer with a classification layer is beneficial, especially when utilizing the smaller versions of the Flan-T5 models.

\noindent \textbf{RQ4 \textit{\rqfour}} Existing SATD identification and classification approaches only take code comments as input and overlook other contextual information. We explore the potential of LLMs to leverage this additional contextual information for improving SATD classification performance. We found that larger fine-tuned models such as Flan-T5-large and Flan-T5-XL can effectively utilize these contextual features to enhance performance. In contrast, smaller models and those employing ICL exhibit a decrease in performance when complex contextual information is included.

Our work makes the following main contributions:
\begin{itemize}
     \item We design and perform a comprehensive evaluation to assess the performance of six popular LLMs in automated SATD identification and classification. This evaluation encompasses various aspects, including different model types, sizes, adaptation methods (fine-tuning vs. ICL), prompting techniques, and model architectures (with or without a classification layer). 
     \item We are the first to investigate the potential of including contextual features beyond code comments in SATD classification task, leveraging LLM. 
     \item We demonstrate that LLMs can achieve state-of-the-art performance in SATD identification and classification. Specifically, a fine-tuned Flan-T5-XL model achieved an F1 score of 0.839 in SATD identification. In SATD classification, the top-performing model was an ensemble fine-tuned Flan-T5-XL, which integrated various combinations of contextual features and code comments, achieving an overall accuracy of 0.668.
     \item The results and source code related to this study are available at \url{https://github.com/RISElabQueens/SATD_LLM}. 
\end{itemize}

The rest of this paper is organized as follows. Section \ref{sec:related-work} summarizes research background and related work. The experimental design and result analysis are presented in Section \ref{sec:study-setup} and \ref{sec:results}, respectively. We discuss the impact of epoch numbers in Section \ref{sec:discussion} and threats to validity in Section \ref{sec:threats}. Finally, the paper concludes with the future work in Section \ref{sec:conclusion}.

\section{Related Work}\label{sec:related-work}
\subsection{Self-Admitted Technical Debt Identification}

The goal of SATD identification is to determine whether a given code comment admitted that the corresponding code is a technical debt or not~\citep{Sheikhaei-2023}. A few studies aim to detect SATD in other software artifacts, such as issues~\citep{Li-2022,Li-2023}, leveraging comment-based SATD approaches. Existing SATD identification approaches can be classified into two categories: 1) rule-based approaches~\citep{Potdar-2014,Guo-2021,Sridharan-2023} which search for certain keywords (e.g., TODO) or phrases (e.g., ``probably a bug'') in the code comment, and 2) supervised learning based approaches~\citep{Huang-2017,Maldonado-2017,Ren-2019,Prenner-2022} which train a model from labeled data and evaluate it on unseen code comments. 

One of the most decent rule-based approaches is Matches task Annotation Tags (MAT)~\citep{Guo-2021}. In this approach, the authors showed that just matching a set of popular task annotation tags, i.e., TODO, FIXME, HACK, and XXX, provides similar or even superior performance for SATD identification compared with traditional machine learning approaches such as the natural language processing (NLP) approach~\citep{Maldonado-2017} which leverages maximum entropy classifier, and the text-mining based approach~\citep{Huang-2017} that employs Naïve Bayes Multinomial classifier. A more recent rule-based approach is PENTACET~\citep{Sridharan-2023}, where the authors extend the 64 SATD identification patterns introduced by Potdar and Shihab~\citep{Potdar-2014}, to 1,041 patterns using a tool named Sense2Vec~\citep{Trask-2015}. Sense2Vec captures contextually similar words with its word embedding and the authors use these extrapolated features (words) to evaluate the code comments for SATD. While it has shown that simple rule-based approaches, e.g., MAT, have a high precision, though not having a good recall, the performance of a complex rule-based approach such as PENTACET is not well studied. In this paper, we consider both MAT and PENTACET approaches as rule-based baselines, and the NLP approach~\citep{Maldonado-2017} as a traditional machine-learning-based baseline for the SATD identification task.

More recently, supervised deep learning-based methods have emerged and achieved more promising performance compared to rule-based and traditional machine-learning approaches. Ren et al.~\citep{Ren-2019} proposed a Convolutional Neural Network (CNN) based approach that outperforms traditional text-mining models. Yu et al.~\citep{Yu-2021} proposed a Bidirectional Long Short-Term Memory (BiLSTM) model with a balanced cross-entropy loss function to overcome the class unbalance challenge. A more recent study by Prenner and Robbes~\citep{Prenner-2022} showed that the BERT base models provide state-of-the-art results in different software-related tasks, including SATD identification. As the CNN and the BERT base models have achieved the state-of-the-art results in this domain, we use CNN as a strong non-LLM baseline, and BERT as a LLM baseline in our experiments for both SATD identification and classification tasks.

\subsection{Self-Admitted Technical Debt Classification}

Maldonado and Shihab~\citep{Maldonado-2015} introduced the task of SATD classification. They extracted 33,093 code comments from five Java projects and manually classified them into six categories: non-SATD, Design, Requirement, Defect, Test, and Documentation. Later, \cite{Maldonado-2017} extended the dataset by adding another five Java projects with the same categories and proposed an NLP approach using the Java implementation of a maximum entropy classifier, Stanford Classifier~\citep{Manning-2003}, to identify and classify SATDs by the code comments. Later, other researchers worked on that dataset and proposed another classification from a different perspective. Fucci et al.~\citep{Fucci-2021} and Cassee et al.~\citep{Cassee-2022} proposed a bottom-up strategy (i.e., what do SATD comments mention?) rather than top-down (i.e., how do SATD comments map into a software development life-cycle?) which was utlized in~\citep{Maldonado-2015}. They manually classified 1,038 SATD instances from Maldonado dataset into 41 categories and subcategories, such as \textit{``poor implemented choices''} and \textit{``functional issues''}. Chen et al.~\citep{Chen-2022} leveraged chi-square to select representative features from the textual features, and applied the XGBoost model for the SATD classification task on the Maldonado dataset.

In addition to SATD classification for general software systems, there are studies that focus on identifying and classifying SATDs in specific domains such as Blockchain projects~\citep{Pinna-2023}, machine learning (ML) software~\citep{OBrien-2022,Bhatia-2023}, and deep learning frameworks~\citep{Liu-2020}. Pinna et al.~\citep{Pinna-2023} employed the NLP approach~\citep{Maldonado-2017} to detect Design SATDs and Requirement SATDs in Blockchain projects. OBrien et al.~\citep{OBrien-2022} manually classified the extracted SATDs into six general categories, which were previously introduced by Bavota and Russo~\citep{Bavota-2016}: Requirement, Code, Test, Defect, Design, and Documentation. Then, they defined 23 specific categories for machine learning-related SATD and manually applied them to their dataset. Bhatia et al.~\citep{Bhatia-2023} employed the text-mining tool~\citep{Liu-2018} to identify SATDs in 318 ML and 318 non-ML open-source software projects. They then manually classified 611 randomly sampled SATDs based on the categories introduced by Bavota and Russo~\citep{Bavota-2016}.

\subsection{Large Language Models}

LLMs are deep learning models~\citep{Goodfellow-2016}, typically based on the Transformer architecture~\citep{Vaswani-2017}, and pre-trained on extensive corpora. One of the early milestones in LLMs was BERT (Bidirectional Encoder Representations from Transformers)~\citep{Devlin-2019}, which pioneered the use of bidirectional training to enhance the understanding of word context within sentences, significantly advancing performance across a range of natural language processing tasks. Following BERT's encoder-only architecture, which excelled mainly in text classification, a new wave of LLMs emerged. These newer models adopted encoder-decoder~\citep{Raffel-2019} or decoder-only architectures~\citep{Radford-2019}, catalyzing a revolution in text generation tasks. Concurrently, variations of the BERT model were introduced, such as CodeBERT~\citep{Feng-2020}, designed for understanding and generating code by blending natural language and programming language training, and RoBERTa~\citep{Liu-2019}, an optimized version of BERT with improved training methodology and larger datasets.

T5 (Text-to-Text Transfer Transformer)~\citep{Raffel-2019} and Flan-T5~\citep{Chung-2022} are notable examples of LLMs utilizing the encoder-decoder architecture. T5, developed by Google, treats every language task as a text-to-text problem, converting tasks like translation, summarization, and question-answering into a unified framework. Flan-T5, an extension of T5, further enhances its capabilities through fine-tuning with a mixture of instruction-based tasks, improving its performance on various benchmarks. Both models are available in five different sizes, ranging from the small size with 77M parameters to the XXL size with 11.1B parameters. This variety enables researchers to study the effect of model size on their applications. Flan-T5 has been employed in various applications such as text summarization~\citep{tam-2023} and log parsing~\citep{Jiang-2023-llmparser} tasks and has demonstrated impressive performance, even when compared to some newer models like LLaMA~\citep{Touvron-2023}.

There are two general approaches for applying LLMs on downstream tasks: 1) \textit{fine-tuning}: continuously train the LLM on the downstream task for a few more epochs, and 2) \textit{in-context learning (ICL)}: instructing the model on what it is expected to do, and if necessary, adding a few examples in the instruction to make the task clearer. The process of finding a good instruction to achieve good results is called \textit{prompt engineering}. The main advantage of ICL over fine-tuning is that it doesn't require investing time and processing resources to update model parameters for the downstream task. In this study, we consider both approaches for the identification and classification of SATDs.

LLMs have been successfully employed in different Software Engineering (SE) tasks. Prenner and Robbes~\citep{Prenner-2022} applied BERT-based models on a selection of 13 smaller datasets from the SE literature, and achieved superior performance for tasks involving natural language. Gao et al.~\citep{Gao-2023} leveraged the OpenAI's Codex and GPT-3.5 models using the ICL approach for code intelligence tasks including code summarization, bug fixing, and program synthesis. These are just two examples in LLM4SE (Large Language Models for Software Engineering) domain. Hou et al.~\citep{Hou-2023} conducted a systematic literature review on LLM4SE by studying 229 research papers from 2017 to 2023. The success of LLMs in various SE-related tasks motivated us to investigate their effectiveness in SATD tasks.

%\section{Research Questions}\label{sec:RQs}
%\input{_RQs}

\section{Study Setup}\label{sec:study-setup}

\subsection{Research Questions}
\paragraph{\textbf{RQ1: \rqone}}
Previous studies have explored rule-based methods, as well as machine learning and deep learning approaches, for identifying and classifying SATD. A notable recent study by Prenner and Robbes~\citep{Prenner-2022} employed the BERT model for several classification tasks in software engineering, including SATD identification, and achieved significant improvements over traditional methods. The success of BERT in this domain, combined with the advent of more sophisticated, larger, open-source language models like T5 and LLaMA, has inspired our investigation of their effectiveness for SATD identification and classification. Therefore, our first research question explores whether these newer and more advanced language models, known for their proficiency in text generation tasks such as summarization and translation, can also outperform BERT in the efficient identification and classification of SATD. In our experiments for RQ1, we employed the Flan-T5 models, BERT, and CodeBERT as the chosen LLMs for fine-tuning and compared their results with non-LLM baselines. The rationale for selecting these models is presented in Section~\ref{subsec:selected_llm}.

\paragraph{\textbf{RQ2: \rqtwo}}
While RQ1 focuses on fine-tuning pre-trained models, this research question explores whether ICL using a larger LLM can surpass the performance of smaller models that have been fine-tuned for SATD identification and classification. ICL is notably cost-effective during the learning phase, just requiring prompt engineering, making it an appealing approach, especially when training data is limited. To assess the efficacy of the ICL approach, we utilize Flan-T5-XXL with 11.1B parameters, which is approximately four times larger than the biggest model fine-tuned in this study, namely, Flan-T5-XL with 2.85B parameters.

\paragraph{\textbf{RQ3: \rqthree}}
In order to employ the Flan-T5 models for SATD identification and classification, as a common practice we refine their architecture by substituting the original text generation layer with a classification layer to better adapt the model for classification tasks (see Section~\ref{subsec:architecture}). However, the impact of this modification is unclear. Furthermore, some studies utilize the original LLM architecture and treat the classification task as a text-to-text problem~\citep{Raffel-2019}, or perform the classification through inference~\citep{Zhang-2023}, specifically by predicting the next token (which is expected to be a class name) when provided with input data as a prompt.
To assess the impact of this architectural modification, we conduct fine-tuning experiments using the original Flan-T5 models, in contrast to RQ1 and RQ4, for which we use the modified architecture. 

\paragraph{\textbf{RQ4: \rqfour}}
In RQ1-RQ3, the models are given only the code comments - the default setting for existing SATD identification and classification approaches. RQ4 aims to explore whether including additional contexts, such as the file path, the containing method's signature, and the body of the containing method, enhances the performance of large language models in the SATD classification task. A key aspect of this investigation is determining the optimal size of the LLM required to utilize this contextual data for improved results effectively. Specifically, we want to assess whether smaller LLMs can discern patterns and relationships between the code comments and the contextual features to enhance SATD classification performance. We investigate this question solely for SATD classification because the largest benchmark for SATD identification, the Maldonado dataset~\citep{Maldonado-2017}, does not provide the location and contextual information for each entry.

\subsection{Datasets}

In this study, we employ two datasets: Maldonado-62k and OBrien. The usage of these datasets is as follows: for RQ1, RQ2, and RQ3, we utilize both datasets. For RQ4, we only use the OBrien dataset, because, unlike the Maldonado-62k dataset, which only provides the comment text for each entry, the OBrien dataset includes additional features such as the file path that can be utilized by LLMs to achieve better performance. More details for these two datasets are provided below.

\paragraph{\textbf{Maldonado-62k:}} Initially, Maldonado et al.~\citep{Maldonado-2015} compiled a SATD dataset by manually labeling 33,093 source code comments from five well-documented open-source projects across various application domains, specifically Apache Ant, Apache JMeter, ArgoUML, Columba, and JFreeChart. This dataset was later expanded in~\citep{Maldonado-2017} to include 62,275 labeled code comments from 10 projects, among which 4,071 were identified as instances of SATD. Subsequently, Yu et al.~\citep{Yu-2022} conducted a meticulous review of all comments initially classified as non-SATD, particularly those containing strong SATD indicators, i.e., todo, fixme, hack, and workaround. They discovered that 426 of the 434 comments, previously labeled as non-SATD, were indeed SATD. As a result, the revised Maldonado dataset comprises 4,497 SATD and 57,778 non-SATD code comments. In our study, we utilize this modified version of the Maldonado dataset, which we refer to as the Maldonado-62k dataset. The dataset comprises three fields: ``project'', ``comment'', and ``label''. However, it lacks contextual details like the surrounding code, commit history, or file path, as these were not included by the dataset creators. Table~\ref{tab:maldonado-62k} shows the statistics of this dataset.

\begin{table}
\centering
\caption{Maldonado-62k dataset statistics}
\label{tab:maldonado-62k} 
\begin{tabular}{lrrr}
\hline\noalign{\smallskip}
Project & \#Comments & \#SATD (original) & \#SATD (Jitterbug) \\
\noalign{\smallskip}\hline\noalign{\smallskip}
ApacheAnt & 4,098 & 131 & 135 \\ 
ArgoUML & 9,452 & 1,413 & 1,630 \\ 
Columba & 6,468 & 204 & 220 \\ 
EMF & 4,390 & 104 & 119 \\ 
Hibernate & 2,968 & 472 & 493 \\ 
JEdit & 10,322 & 256 & 259 \\ 
JFreeChart & 4,408 & 209 & 247 \\ 
JMeter & 8,057 & 374 & 416 \\ 
JRuby & 4,897 & 622 & 665 \\ 
SQuirrel & 7,215 & 286 & 313 \\ 
\noalign{\smallskip}\hline\noalign{\smallskip}
Total & 62,275 & 4,071 & 4,497 \\ 
\noalign{\smallskip}\hline
\end{tabular}
\end{table}

\paragraph{\textbf{OBrien:}} OBrien et al.~\citep{OBrien-2022} compiled a dataset from popular machine learning (ML) repositories written in Python, including scikit-learn and IntelPython. This dataset comprises 856 labeled instances of SATD identified within these repositories. The labeling process introduced various layers, such as ``Software TD Type'', ``ML TD Type'', and ``ML Pipeline Stage''. However, our study focuses on the general classification of SATD, so we selected only the `Software TD Type' column as our target label, omitting the others. We refined this dataset by eliminating entries with null values in the ``Software TD Type'' column, resulting in a total of 789 entries. The labels' descriptions and frequencies are detailed in Table~\ref{tab:obrien-dataset}. A notable feature of this dataset is the inclusion of contextual information for each entry, such as the file path and the commit introducing the SATD. Leveraging this data, we added two contextual columns: ``containing method signature'' and ``containing method body''. Among 789 entries in this dataset, 447 SATD instances are located either within a method or on the very first line preceding a method, allowing us to associate each with its corresponding method.

There are two primary reasons for selecting the OBrien dataset for the SATD classification task. First, we required a dataset comprising self-admitted technical debts introduced in code comments. This is to access more advanced data, such as the surrounding code, which is crucial for delivering more accurate predictions of SATD types. Therefore, we are not considering datasets focused on other sources like issue trackers~\citep{Xavier-2020} or build systems~\citep{Xiao-2021}. Second, among the datasets that categorize SATD in code comments, only the OBrien dataset~\citep{OBrien-2022} offers additional information, namely the file path and commit hash. This information is essential for accessing the contextual details for each code comment. In contrast, the Maldonado datasets~\citep{Maldonado-2015,Maldonado-2017} and other datasets, such as~\citep{Fucci-2021,Cassee-2022}, which are derived from the Maldonado dataset, lack such data, making it challenging to extract the surrounding code for each code comment.

\begin{table}
\caption{Definitions and examples of the six SATD types identified in OBrien dataset~\citep{OBrien-2022}}
\label{tab:obrien-dataset} 
\resizebox{12cm}{!}{
\begin{tabular}{|p{0.16\linewidth}|p{0.40\linewidth}|p{0.34\linewidth}|p{0.1\linewidth}|}
\hline
SATD Types & Description & Example Comment & \#of Occ. \\
\hline
Requirement & Requirement debts can be functional or non-functional. In the functional case, implementations are left unfinished or in need of future feature support. In the non-functional case, the corresponding code does not meet the requirement standards (speed, memory usage, ecurity, etc...). & TODO: handle channel modalities later, TODO: make efficient, TODO: Implement Conv Transpose. & 321 \\
\hline
Code & Bad coding practices leading to poor legibility of code, making it difficult to understand and maintain. & TODO: This next code is dense and confusing. Clean up at some point. & 207 \\
\hline
Test & Problems found in implementations involving testing or monitoring subcomponents. & XXX: should we rather test if instance of estimator? & 84 \\
\hline
Defect & Identified defects in the system that should be addressed. & TODO this will fail if a parameter cant handle size=(N;) & 82 \\
\hline
Design & Areas which violate good software design practices, causing poor flexibility to evolving business needs. & TODO maybe improve this so it doesn’t use a global & 80 \\
\hline
Documentation & Inadequate documentation that exists within the software system. & TODO update doc above & 15 \\
\hline
\end{tabular}
}
\end{table}

Both datasets present challenges. In the Maldonado-62k dataset, out of 62,275 code comments, only 4,497 items (7.2\%) are classified as SATD. In the OBrien dataset, there are only 789 items. Therefore, these two datasets also challenge the effectiveness of large language models in dealing with imbalanced data (for SATD identification) and the scarcity of labeled data (for SATD classification).

\subsection{Selected Large Language Models}\label{subsec:selected_llm}

For RQ1 and RQ4, we consider the following models:

\begin{itemize}
    \item \textbf{BERT-base-uncased} (110 million parameters)~\citep{Devlin-2019}: This model has demonstrated superior performance in the SATD identification task~\citep{Prenner-2022}, establishing a strong baseline for other LLMs in SATD identification and classification tasks.
    \item \textbf{CodeBERT} (125 million parameters)~\citep{Feng-2020}: Differing from BERT, which primarily focuses on natural language, CodeBERT is a bimodal extension that utilizes both natural language and source codes. This makes it particularly suited for processing code comment SATD.
    \item \textbf{Flan-T5}~\citep{Chung-2022}: This is an enhanced version of T5~\citep{Raffel-2019}, fine-tuned on over 1000 tasks. Available in various sizes, Flan-T5 allows for an exploration of the impact of model size on SATD tasks. Our infrastructure supports full fine-tuning of all Flan-T5 variants except the XXL version (11.1 billion parameters). Therefore, we will include Flan-T5 Small (77 million parameters), Base (248 million parameters), Large (783 million parameters), and XL (2.85 billion parameters) in our experiments.
\end{itemize}

For RQ2, we will utilize Flan-T5-XXL, the largest variant with 11.1 billion parameters. This model does not require fine-tuning for our purposes, as we will employ the ICL approach for the SATD identification and classification tasks. For RQ3, we evaluate the performance of the original Flan-T5 architecture against their altered version utilized in RQ1 and RQ4.

\subsection{Baselines} 
For the task of SATD identification, we employ four baseline methods: the NLP approach by~\citep{Maldonado-2017}, the MAT method~\citep{Guo-2021}, the technique introduced in the PENTACET study~\citep{Sridharan-2023}, and the convolutional neural networks (CNN) model proposed by~\citep{Ren-2019}. For the SATD classification task, we utilize two baselines: the NLP approach~\citep{Maldonado-2017} and the CNN model~\citep{Ren-2019}.

\subsection{Evaluation metrics}
Following the previous studies~\citep{Prenner-2022,Ren-2019,Maldonado-2017}, we report the F1 score for the SATD class in the SATD identification task. For the SATD classification task, we report overall performance using accuracy, and per class performance using the F1 score.

\subsection{Updating Flan-T5 architecture for the classification task}\label{subsec:architecture}

\begin{figure*}
\includegraphics[width=1.0\textwidth]{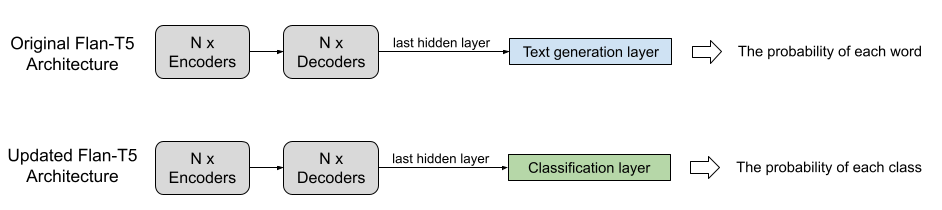}
\caption{Updating Flan-T5 architecture by replacing the last layer with a classification layer}
\label{fig:updating-flan-t5-architecture}
\end{figure*}

In this study, we propose an adaptation of the Flan-T5 architecture to address the classification task (see Figure~\ref{fig:updating-flan-t5-architecture}). The original Flan-T5 model is inherently designed for text generation tasks, where the goal is to predict the probability distribution of subsequent words in a given text sequence. This is achieved through the model's final layer, which is a text generation layer that outputs probabilities for each word in the vocabulary.

To tailor this architecture to classification objectives, we replace the text generation layer with a classification head, which consists of a dropout layer followed by a linear neural network layer with the size of number of classes. This new layer interprets the semantic representation of the input sequence from the last hidden state and produces a probability distribution over the predefined classes. For the loss function, we use cross-entropy which is a common choice for classification tasks. This modification leverages the pre-existing language understanding capabilities of the Flan-T5 model while aligning the output structure with the requirements of classification tasks. The updated architecture now effectively assigns input sequences to categories. We use this architecture in RQ1 and RQ4. For RQ2, we use the original architecture because for ICL approaches, there is no need to update the model parameters. In RQ3, we evaluate the effectiveness of the modified architecture against the original one.

\subsection{Implementation}
We begin by downloading the official checkpoints for all evaluated models from Hugging Face. Subsequently, the Torch and Transformers packages are employed to conduct fine-tuning and ICL on a single Nvidia RTX 6000 GPU with 48 GB VRAM. In all models, we use the default 32-bit precision mode, and all experiments are done in Python 3.10.

For all baseline models, we employed their default settings. For selected LLMs, we adjusted the batch size to the highest power-of-two value that our VRAM could support. Additionally, since the learning rate significantly influences LLM performance, we conducted a few experiments to determine a near-optimal learning rate for each LLM. In the CNN and all LLMs, we configured the maximum length of input data for the Maldonado-62k dataset at 128 tokens. For the OBrien dataset, we set it at 512 tokens. The rationale for the higher token limit in the OBrien dataset is related to RQ4, where we explore the impact of incorporating contextual data into the input. This necessitates additional space to include this information for the models. All large language models are trained for eight epochs using the AdamW~\citep{Loshchilov-2019} optimizer with a linear learning rate decay schedule. As there is no validation set to choose the best model across training epochs, we take the resulting model of the last epoch and report its performance on the test data. Table~\ref{tab:llm-training-parameters} presents the key parameter settings for each model applied to each dataset.

\begin{table}
\centering
\caption{The key training parameter settings for each LLM applied to each dataset. All models are trained for 8 epochs.}
\label{tab:llm-training-parameters} 
\begin{tabular}{lllll}
\hline\noalign{\smallskip}
\textbf{Model} & \textbf{Dataset} & \textbf{Maxlen} & \textbf{Batch} & \textbf{Learning} \\
               &                  &                 & \textbf{size}  & \textbf{rate} \\
\noalign{\smallskip}\hline\noalign{\smallskip}
BERT-base-uncased (110M) & Maldonado-62k & 128 & 32 & 0.00001 \\ 
CodeBERT-base (125M) & Maldonado-62k & 128 & 32 & 0.00001 \\ 
Flan-T5-small (77M) & Maldonado-62k & 128 & 32 & 0.0001 \\ 
Flan-T5-base (248M) & Maldonado-62k & 128 & 32 & 0.0001 \\ 
Flan-T5-large (783M) & Maldonado-62k & 128 & 16 & 0.0001 \\ 
Flan-T5-XL (2.85B) & Maldonado-62k & 128 & 4 & 0.00002 \\ 
\noalign{\smallskip}\hline\noalign{\smallskip}
BERT-base-uncased (110M) & OBrien & 512 & 32 & 0.00005 \\ 
CodeBERT-base (125M) & OBrien & 512 & 32 & 0.00005 \\ 
Flan-T5-small (77M) & OBrien & 512 & 32 & 0.001 \\ 
Flan-T5-base (248M) & OBrien & 512 & 16 & 0.0005 \\ 
Flan-T5-large (783M) & OBrien & 512 & 4 & 0.0002 \\ 
Flan-T5-XL (2.85B) & OBrien & 512 & 1 & 0.00005 \\ 
\noalign{\smallskip}\hline
\end{tabular}
\end{table}

\section{Results and Analysis}\label{sec:results}
\subsection{\rqone}\label{subsec:results_RQ1}
\subsubsection{Approach}

\paragraph{\noindent\textbf{SATD Identification.}} The rule-based baselines, MAT and PENTACET, do not require training data. Therefore, they are run on the entire dataset at once. In contrast, other baselines and the LLMs necessitate training data. In line with previous studies~\citep{Prenner-2022,Ren-2019}, a cross-project approach is employed to evaluate these models. As the Maldonado dataset contains 10 projects, each round involves selecting one project as the test project and using the remaining nine for model training.

\paragraph{\noindent\textbf{SATD Classification.}} A similar method is applied to the OBrien dataset, but with 10 randomly assigned folds instead of cross-project validation. In each round, one fold is used for testing, and the remaining nine for training. The data is split into 10 folds only once, and this division is used across all experiments on this dataset. Each fold comprises 79 data items, except the last one, which contains 78. Given the relatively small size of the OBrien dataset, LLM models may show varying performance in different runs. Therefore, each experiment is conducted three times with different seeds, and the average accuracy or F1 score is reported.

\subsubsection{Results}
\paragraph{\textbf{SATD Identification.}} Table~\ref{tab:maldonado62k-baselines-vs-LLMs} shows the F1 score of all selected large language models for each of the 10 Java projects against the four baselines, NLP~\citep{Maldonado-2017}, CNN~\citep{Ren-2019}, MAT~\citep{Guo-2021}, and PENTACET~\citep{Sridharan-2023}. The final row, which presents the average F1 score across the 10 projects, demonstrates that all six large language models significantly outperform the four baselines, with margins ranging from 4.4\% to 11.3\%. Moreover, in terms of F1 score per project, the top-performing model is an LLM for all projects except Ruby, where MAT achieves an F1 score of 95.1\%, marginally higher (by 0.1\%) than the scores of Flan-T5 small and large models. Among all models, Flan-T5-XL delivers the highest average score, despite not achieving the best score in every project. The results from the large language models further suggest that larger models generally surpass their smaller counterparts, supporting the notion that larger models tend to yield better results.

The Flan-T5-XL model is 37 times larger than the smallest model, Flan-T5-small, but only slightly outperforms it, with an F1 score improvement of just 2.1\% (83.9\% vs 81.8\%). This modest improvement is surprising considering the model's size. One significant challenge in model performance is the quality of annotations. To delve deeper, we analyzed Cohen’s Kappa coefficient, comparing Flan-T5-XL's predictions with the ground truth. The coefficient is 0.89, higher than the 0.81 inter-rater agreement reported in the study, indicating that our Flan-T5-XL model is more accurate than human annotators. This outcome may appear counter-intuitive, as the model was trained on imperfect human labels. However, Flan-T5-XL is not a simple model; it's extensively pre-trained on a large text corpus, enabling it to understand the meaning of words and phrases. The capability, obtained by combining its pre-training and fine-tuning, likely helps it to identify and ignore obvious mislabeling and make more accurate predictions. Indeed, improving the quality of annotated data is a prerequisite for further performance improvement.

\begin{table}
\centering
\caption{Comparative F1 scores of six large language models and baselines in SATD identification on Maldonado-62k dataset. The ``cross-project'' row indicates if we employ cross-project prediction, i.e., using 9 projects for training and the remaining 1 for testing.}
\label{tab:maldonado62k-baselines-vs-LLMs}
\resizebox{12cm}{!}{
\begin{tabular}{l|llll|llllll}
\hline\noalign{\smallskip}
 & NLP & MAT & PENT- & CNN & BERT & CodeBERT & Flan-T5 & Flan-T5 & Flan-T5 & Flan-T5 \\
 &     &     & ACET  &     & base & base     & small   & base    & large   & XL \\
Cross-project & yes & no & no & yes & yes & yes & yes & yes & yes & yes \\ 
\noalign{\smallskip}\hline\noalign{\smallskip}
ApacheAnt & 0.532 & 0.654 & 0.505 & 0.625 & 0.629 & 0.691 & 0.643 & 0.679 & 0.689 & \textbf{0.710} \\ 
ArgoUML & 0.936 & 0.934 & 0.894 & 0.946 & 0.954 & 0.954 & 0.956 & 0.951 & \textbf{0.958} & 0.939 \\ 
Columba & 0.833 & 0.928 & 0.892 & 0.88 & 0.857 & 0.941 & 0.923 & 0.919 & \textbf{0.944} & 0.933 \\ 
EMF & 0.527 & 0.434 & 0.574 & 0.420 & \textbf{0.746} & 0.651 & 0.660 & 0.693 & 0.707 & 0.733 \\ 
Hibernate & 0.852 & 0.851 & 0.859 & 0.892 & 0.902 & 0.893 & 0.895 & 0.901 & 0.913 & \textbf{0.922} \\ 
JEdit & 0.470 & 0.321 & 0.654 & 0.605 & 0.631 & 0.668 & 0.688 & 0.698 & \textbf{0.727} & 0.717 \\ 
JFreeChart & 0.705 & 0.707 & 0.704 & 0.739 & 0.734 & 0.738 & \textbf{0.746} & 0.735 & 0.741 & 0.734 \\ 
JMeter & 0.819 & 0.870 & 0.795 & 0.868 & 0.879 & 0.878 & 0.874 & \textbf{0.883} & \textbf{0.883} & 0.882 \\ 
JRuby & 0.896 & \textbf{0.951} & 0.928 & 0.919 & 0.947 & 0.940 & 0.950 & 0.948 & 0.950 & 0.945 \\ 
SQuirrel & 0.689 & 0.731 & 0.776 & 0.777 & 0.826 & 0.842 & 0.846 & 0.834 & 0.834 & \textbf{0.869} \\ 
\noalign{\smallskip}\hline\noalign{\smallskip}
Average & 0.726 & 0.738 & 0.758 & 0.767 & 0.811 & 0.820 & 0.818 & 0.824 & 0.835 & \textbf{0.839} \\
\noalign{\smallskip}\hline
\end{tabular}
}
\end{table}

\paragraph{\noindent\textbf{SATD Classification.}} Table~\ref{tab:obrien-f1-per-category} presents the per-category F1 scores and overall accuracy for six LLMs and two baselines, NLP and CNN, for the SATD classification task. In contrast to the SATD identification task, where smaller models like Flan-T5-small and BERT-base performed competitively with larger models, a more pronounced difference is observed in this table. Here, the larger models significantly outperform the smaller ones. The Flan-T5-XL achieves the highest accuracy at 0.62, while the Flan-T5-small registers the lowest at 0.537. This disparity in performance can be attributed to two main factors. Firstly, the training dataset for SATD identification task consists of approximately 50,000 to 60,000 items, varying based on the selected project as the test data. In comparison, the training dataset for SATD classification is considerably smaller, with only 710 items. Secondly, the nature of the tasks differs: SATD identification is just a binary classification, whereas SATD classification requires a more complex 6-class classification. This additional complexity poses a greater challenge, particularly for smaller models. Consequently, the larger pre-trained models, which exhibit a more profound understanding of textual data, demonstrate superior efficacy in learning and performing the classification task compared to their smaller counterparts.

Table~\ref{tab:obrien-f1-per-category} also shows that the performance of NLP is lower than even the smallest LLM, while CNN achieves a commendable overall accuracy of 0.602, close to the best-performing LLM, Flan-T5-XL, which has an accuracy of 0.620. This effectiveness of CNN led to its selection as a strong baseline for our last research question.

Among six categories in the OBrien dataset, Flan-T5-XL scores the highest F1 in three categories and the second-highest in two of the remaining three categories. A noteworthy observation is the ``Document" class, containing only 15 items, where all LLMs show low F1 scores, ranging between 0.0 and 0.23. However, Flan-T5-XL distinguishes itself with a relatively high F1 score of 0.488 in this class, indicative of its consistent performance across varied class sizes. While CNN performs best in the ``Document" class, its performance is not uniformly high across all categories. For instance, it scored second-worst in the ``Defect" category.

\begin{table}
\caption{Average F1 score for each category and the overall accuracy (Dataset: OBrien, Approach: 10-fold cross validation, number of runs: 3)}
\label{tab:obrien-f1-per-category}
\resizebox{12cm}{!}{
\begin{tabular}{llllllll}
\hline\noalign{\smallskip}
Class label & Code & Defect & Design & Doc. & M\&T & Requirement & All \\ 
Count & 207 & 82 & 80 & 15 & 84 & 321 & 789 \\ 
\noalign{\smallskip}\hline\noalign{\smallskip}
NLP & 0.512 & 0.290 & 0.256 & 0.300 & 0.549 & 0.632 & 0.527 \\ 
CNN & 0.627 & 0.247 & 0.357 & \textbf{0.545} & 0.611 & \textbf{0.692} & 0.602 \\ 
BERT-base-uncased (110M) & 0.585 & 0.367 & 0.312 & 0.170 & \textbf{0.634} & 0.662 & 0.576 \\ 
CodeBERT-base (125M) & 0.636 & 0.398 & 0.423 & 0.176 & 0.610 & 0.687 & 0.611 \\ 
Flan-T5-small (77M) & 0.533 & 0.120 & 0.074 & 0.0 & 0.600 & 0.659 & 0.537 \\ 
Flan-T5-base (248M) & 0.600 & 0.378 & 0.299 & 0.0 & 0.593 & 0.654 & 0.565 \\ 
Flan-T5-large (783M) & 0.586 & 0.418 & 0.329 & 0.230 & 0.598 & 0.662 & 0.575 \\ 
Flan-T5-XL (2.85B) & \textbf{0.651} & \textbf{0.433} & \textbf{0.475} & 0.488 & 0.618 & 0.684 & \textbf{0.620} \\ 
\noalign{\smallskip}\hline
\end{tabular}
}
\end{table}

\paragraph{\textbf{Summary:}} In SATD identification, the selected LLMs outperformed all existing baselines, with Flan-T5-XL achieving the highest F1 score, a 7.2\% improvement compared to the best-performing baseline. In the SATD classification task, while the fine-tuned Flan-T5-XL still led in performance, the CNN model exhibited competitive results, even surpassing four of six LLMs. In both tasks, the larger models outperformed their smaller counterparts.

%%%%%%%%%%%%%%%%%%%%%%%%%%%%%%%%%%%%%%%%%%%%%%%%%%%%%%%%%%%%%%%%%%%%%%

\subsection{\rqtwo}

\subsubsection{Approach}
In both tasks, we apply the ICL method to the largest model in the Flan-T5 family, Flan-T5-XXL, and compare its performance with that of the fine-tuned smaller versions from the same family. To perform the inference, we set $max\_new\_tokens=5$ and $temperature=0.0$. If none of the class names appear in the model's response (indicating unknown labels), the majority class, i.e., ``non-SATD" for Maldonado-62k dataset and ``Requirement" for OBrien dataset, is assumed to be the predicted class.

\paragraph{\textbf{SATD Identification.}} For SATD identification, we start with employing four distinct zero-shot prompts, each providing a definition of SATD and incorporating various keywords indicative of SATD, such as TODO and FIXME. This approach is based on findings by~\cite {Guo-2021} and~\cite{Yu-2022}, which highlight the significance of these keywords in classifying a code comment as SATD. The prompt for each selected group of keywords is shown in Table~\ref{tab:maldonado62k-icl-prompts}. Once we find the best performing zero-shot prompt, we investigate the effect of incorporating different numbers of examples (1, 2, 3, 5, 10, 15, 20) in that prompt. Two methods are used to select samples for prompts: 1) Randomly choosing $n$ examples from the training projects for each item in the test project, and 2) Using SentenceTransformer (with all-MiniLM-L6-v2 model) to embed the input data and applying cosine similarity to select the most relevant items from the training projects for each item in the test project. 

\begin{table}
\caption{The prompts applied in ICL approach on Maldonado-62k dataset}
\label{tab:maldonado62k-icl-prompts} 
\begin{tabular}{|p{0.15\linewidth}|p{0.78\linewidth}|}
\hline
\textbf{Approach} & \textbf{Prompt} \\
\hline
No keywords & \texttt{Self-admitted technical debt (SATD) is technical debt admitted by the developer through source code comments. Assign the label of SATD or Not-SATD for each given source code comment.} \\ 
\hline
Suggested keywords by MAT & \texttt{Self-admitted technical debt (SATD) are technical debt admitted by the developer through source code comments. SATD comments usually contain specific keywords: TODO, FIXME, HACK, and XXX. Assign the label of SATD or Not-SATD for each given source code comment.} \\ 
\hline
Suggested keywords by Jitterbug & \texttt{Self-admitted technical debt (SATD) is technical debt admitted by the developer through source code comments. SATD comments usually contain specific keywords: TODO, FIXME, HACK, and Workaround. Assign the label of SATD or Not-SATD for each given source code comment.} \\ 
\hline
Suggested keywords by GPT4 & \texttt{Self-admitted technical debt (SATD) is technical debt admitted by the developer through source code comments. SATD comments usually contain specific keywords: TODO, FIXME, HACK, XXX, NOTE, DEBT, REFACTOR, OPTIMIZE, TEMP, WORKAROUND, KLUDGE, REVIEW, NOFIX, PENDING, and BUG. Assign the label of SATD or Not-SATD for each given source code comment.} \\ 
\hline
\end{tabular}
\end{table}

\paragraph{\textbf{SATD Classification.}} For SATD classification, we explore the impact of varying the number of examples (0, 1, 2, 3, 5, 10, 15, 20) in the prompt on ICL performance. Same as the SATD identification approach, two methods are used to select samples for prompts: 1) Randomly choosing n examples from the training folds for each item in the test fold, and 2) Using SentenceTransformer (with all-MiniLM-L6-v2 model) to embed the input data and applying cosine similarity to select the most relevant items from the training folds for each test fold item. Additionally, we explore whether including category descriptions in the prompt enhances classification performance. Table~\ref{tab:obrien-sample-one-shot-prompts} presents a sample one-shot prompt for each prompting approach. 

\begin{table}
\caption{Sample one-shot prompts for SATD classification task}
\label{tab:obrien-sample-one-shot-prompts} 
\resizebox{12cm}{!}{
\begin{tabular}{|p{0.23\linewidth}|p{0.77\linewidth}|}
\hline
\textbf{Prompt approach} & \textbf{Sample one-shot prompt} \\
\hline
Describing categories & \texttt{There are six types of software technical debts:} \\
 + &  \\
The most relevant examples using SentenceTransformer & \texttt{Requirement: Requirement debts can be functional or non-functional. In the functional case, implementations are left unfinished or in need of future feature support. In the non-functional case, the corresponding code does not meet the requirement standards (speed, memory usage, security, etc...).} \\[1ex]
 & \texttt{Code: Bad coding practices lead to poor legibility of code, making it difficult to understand and maintain.} \\[1ex]
 & \texttt{M\&T: Problems found in implementations involving testing or monitoring subcomponents.} \\[1ex]
 & \texttt{Defect: Identified defects in the system that should be addressed.} \\[1ex]
 & \texttt{Design: Areas which violate good software design practices, causing poor flexibility to evolving business needs.} \\[1ex]
 & \texttt{Documentation: Inadequate documentation that exists within the software system.} \\[2ex]
 & \texttt{Here are some examples:} \\[1ex]
 % & \texttt{\#\#\# file path: python/basic\_fusion.py} \\
 & \texttt{\#\#\# Technical debt comment: """ TODO normalize to make sum up to 1? """} \\
 & \texttt{\#\#\# Label: Requirement} \\[1ex]
 % & \texttt{\#\#\# file path: test/torch\_test.py} \\
 & \texttt{\#\#\# Technical debt comment: """ self.mpc\_sum(3; -5) TODO: Future work: how to handle gracefully minus numbers """} \\
 & \texttt{\#\#\# Label:} \\
\hline
Describing categories & \texttt{There are six types of software technical debts:} \\
 + &  \\
Some random examples & \texttt{[describing categories the same as in the above prompt]} \\[2ex]
 & \texttt{Here are some examples:} \\[1ex]
 % & \texttt{\#\#\# file path: file path: textrank/textrank\_word.py} \\
 & \texttt{\#\#\# Technical debt comment: """ TODO: delete this method when no longer needed """} \\
 & \texttt{\#\#\# Label: Code} \\[1ex]
 % & \texttt{\#\#\# file path: test/torch\_test.py} \\
 & \texttt{\#\#\# Technical debt comment: """ self.mpc\_sum(3; -5) TODO: Future work: how to handle gracefully minus numbers """} \\
 & \texttt{\#\#\# Label:} \\
\hline
% Just the most & \texttt{\#\#\# file path: python/basic\_fusion.py} \\
relevant examples using & \texttt{\#\#\# Technical debt comment: """ TODO normalize to make sum up to 1? """} \\
SentenceTransformer & \texttt{\#\#\# Label: Requirement} \\[1ex]
 % & \texttt{\#\#\# file path: test/torch\_test.py} \\
 & \texttt{\#\#\# Technical debt comment: """ self.mpc\_sum(3; -5) TODO: Future work: how to handle gracefully minus numbers """} \\
 & \texttt{\#\#\# Label:} \\
\hline
\end{tabular}
}
\end{table}

\subsubsection{Results}
\paragraph{\textbf{SATD Identification.}} The result for each zero-shot prompting approach is shown in Table~\ref{tab:maldonado62k-zero-shot-f1}. The highest precision and F1 score were achieved by the prompt incorporating keywords from the MAT approach~\citep{Guo-2021}, while the best recall was obtained without including any keywords. The MAT keyword-based prompt attained an F1 score of 0.747, exceeding the results of the NLP and MAT approaches, but falling short of the CNN approach and significantly lagging behind all fine-tuned LLMs (see Table~\ref{tab:maldonado62k-baselines-vs-LLMs}). Table~\ref{tab:maldonado62k-few-shots-f1} presents the F1 scores for the few-shot approach when the MAT keywords are incorporated into the prompt. The table also shows the count of unknown labels (instances where the model's generated tokens do not correspond to any label), noted in parentheses beneath each F1 score. Notably, the few-shot method, whether using random samples or relevant examples by SentenceTransformer, performs worse than the zero-shot approach. This is likely due to data imbalance and the inadequacy of methods in selecting suitable examples for the prompt in this task. According to these results, in the context of SATD identification, the most effective ICL method, which is the zero-shot prompt with MAT keywords, does not outperform even the smallest fine-tuned Flan-T5 model, despite utilizing the largest Flan-T5 model.

\begin{table}
\centering
\caption{Average precision, recall, and F1 score over 10 projects, obtained by the zero-shot ICL approach using Flan-T5-XXL (11.1B) on Maldonado-62k dataset}
\label{tab:maldonado62k-zero-shot-f1} 
\begin{tabular}{llll}
\hline\noalign{\smallskip}
\textbf{Prompt approach} & \textbf{P} & \textbf{R} & \textbf{F1} \\
\noalign{\smallskip}\hline\noalign{\smallskip}
Zero-shot prompt with no keywords & 0.212 & \textbf{0.889} & 0.329 \\ 
Zero-shot prompt with suggested keywords by MAT & \textbf{0.741} & 0.762 & \textbf{0.747} \\ 
Zero-shot prompt with suggested keywords by Jitterbug & 0.678 & 0.76 & 0.712 \\ 
Zero-shot prompt with suggested keywords by GPT4 & 0.585 & 0.791 & 0.664 \\ 
\noalign{\smallskip}\hline
\end{tabular}
\end{table}

\begin{table}
\centering
\caption{Average F1 score over 10 projects, obtained by the few-shot ICL approach using Flan-T5-XXL (11.1B) on Maldonado-62k dataset. Note: Counts of unknown labels (instances where generated tokens do not match any label) are shown in parentheses below each F1 score.}
\label{tab:maldonado62k-few-shots-f1} 
\resizebox{12cm}{!}{
\begin{tabular}{ccccccccc}
\hline\noalign{\smallskip}
\textbf{Prompt approach} & \multicolumn{8}{c}{\textbf{Number of examples in the prompt}} \\ 
 & 0 & 1 & 2 & 3 & 5 & 10 & 15 & 20 \\ 
\noalign{\smallskip}\hline\noalign{\smallskip}
Mentioning MAT keywords + & \textbf{0.747} & 0.687 & 0.673 & 0.661 & 0.662 & 0.653 & 0.650 & 0.648 \\ 
Some examples by & (0) & (0) & (0) & (0) & (0) & (0) & (0) & (0) \\ 
SentenceTransformer & & & & & & & & \\[1ex]
Mentioning MAT keywords + & \textbf{0.747} & 0.707 & 0.679 & 0.665 & 0.660 & 0.649 &  0.651 & 0.649 \\ 
Some random examples   & (0) & (0) & (0) & (6) & (6) & (6) & (6) & (6) \\[2ex]
\noalign{\smallskip}\hline
\end{tabular}
}
\end{table}

\paragraph{\textbf{SATD Classification.}} Table~\ref{tab:obrien-icl-accuracy} displays the accuracy along with the count of unknown labels noted in parentheses beneath each accuracy value. According to the table, unknown labels tend to occur when no examples or category definitions are provided in the prompt.

Key findings from Table~\ref{tab:obrien-icl-accuracy} include:

\begin{itemize}
    \item Providing relevant examples consistently yields better performance compared to random examples.
    \item Including category descriptions before examples in the prompt leads to significantly improved performance, regardless of the number of examples provided in the prompt.
    \item Flan-T5-XXL's ICL approach does not outperform fine-tuning Flan-T5-XL or Flan-T5-large, though it does surpass the performance of fine-tuned Flan-T5-base and Flan-T5-small.
\end{itemize}

\paragraph{\textbf{Summary:}} In the ICL approach using the Flan-T5-XXL model for the SATD identification task, the zero-shot approach provides competitive results with traditional approaches, but performs 6.4\% to 9.2\% worse than fine-tuned LLMs. In contrast, for the SATD classification task, incorporating relevant examples and category descriptions into the prompts enhances performance, surpassing the results achieved by fine-tuning the Flan-T5 small and base models. However, it falls behind the results obtained by fine-tuning the Flan-T5 large and XL models.

\begin{table}
\centering
\caption{Average accuracy over 10 folds, obtained by the ICL approach using Flan-T5-XXL (11.1B) on the OBrien dataset. Note: Counts of unknown labels (instances where generated tokens do not match any label) are shown in parentheses below each F1 score.}
\label{tab:obrien-icl-accuracy} 
\resizebox{12cm}{!}{
\begin{tabular}{ccccccccc}
\hline\noalign{\smallskip}
\textbf{Prompt approach} & \multicolumn{8}{c}{\textbf{Number of examples in the prompt}} \\ 
 & 0 & 1 & 2 & 3 & 5 & 10 & 15 & 20 \\ 
\noalign{\smallskip}\hline\noalign{\smallskip}
Describing categories + & 0.506 & 0.530 & 0.560 & 0.556 & 0.567 & \underline{\textbf{0.572}} & 0.561 & 0.536 \\ 
Some examples by & (12) & (0) & (0) & (0) & (0) & (0) & (0) & (0) \\ 
SentenceTransformer & & & & & & & & \\[1ex]
Describing categories + & 0.506 & 0.511 & 0.513 & \underline{0.521} & 0.518 & 0.510 & 0.493 & 0.450 \\ 
Some random examples   & (12) & (0) & (0) & (0) & (0) & (0) & (0) & (0) \\[2ex]
Just use SentenceTransformer & - & 0.456 & 0.502 & \underline{0.520} & 0.516 & 0.477 & 0.496 & 0.470 \\ 
to mention some examples     &   & (168) & (93) & (45) & (9) & (0) & (1) & (0) \\ 
\noalign{\smallskip}\hline
\end{tabular}
}
\end{table}

%%%%%%%%%%%%%%%%%%%%%%%%%%%%%%%%%%%%%%%%%%%%%%%%%%%%%%%%%%%%%%%%%%%%%%

\subsection{\rqthree}

\subsubsection{Approach}

\paragraph{\noindent\textbf{SATD Identification.}} As described in Section~\ref{subsec:architecture}, we have refined the architecture of the Flan-T5 model by substituting its original text generation layer with a classification layer to better adapt the model for classification tasks. To assess the impact of this architectural modification, we conducted some experiments using the original Flan-T5 model, and inputting only the comment text. To perform the inference, similar to RQ2 approach, we set $max\_new\_tokens=5$ and $temperature=0.0$. If none of the class names appear in the model's response, the majority class is assumed to be the predicted class.

\paragraph{\noindent\textbf{SATD Classification.}} Same as above.

\subsubsection{Results}
\paragraph{\textbf{SATD Identification.}} Table~\ref{tab:maldonado62k-original-vs-altered-architecture} presents the comparison between the original model and the modified version with the added classification layer. We observe that the two architectures achieve nearly identical performance.

\paragraph{\textbf{SATD Classification.}} Table~\ref{tab:obrien-original-vs-altered-architecture} presents the comparison between the original model and the modified version for the classification task. In contrast to the outcomes observed for the identification task, the modified architecture demonstrates significant improvement over the original in the classification task, particularly when employing the smaller versions of the Flan-T5 models. We believe this enhancement can be attributed to the lesser amount of available training data for the classification task, compared to the identification task.

\begin{table}
\centering
\caption{Comparison of average F1 score: original vs. modified Flan-T5 architecture on the Maldonado-62k dataset using comment text as input}
\label{tab:maldonado62k-original-vs-altered-architecture} 
\begin{tabular}{lll}
\hline\noalign{\smallskip}
Model & Original architecture & Adding the classification layer \\
\noalign{\smallskip}\hline\noalign{\smallskip}
Flan-T5-small & 0.820 & 0.818 \\ 
Flan-T5-base & 0.820 & 0.824 \\ 
Flan-T5-large & 0.835 & 0.835 \\ 
Flan-T5-XL &  0.831 & 0.839 \\ 
\noalign{\smallskip}\hline
\end{tabular}
\end{table}

\begin{table}
\centering
\caption{Comparison of average accuracy: original vs. modified Flan-T5 architecture on the OBrien dataset using comment text as input}
\label{tab:obrien-original-vs-altered-architecture} 
\begin{tabular}{lll}
\hline\noalign{\smallskip}
Model & Original architecture & Adding the classification layer \\
\noalign{\smallskip}\hline\noalign{\smallskip}
Flan-T5-small & 0.447 & 0.537 \\ 
Flan-T5-base & 0.505 & 0.565 \\ 
Flan-T5-large & 0.561 & 0.575 \\ 
Flan-T5-XL & 0.602 & 0.620 \\  
\noalign{\smallskip}\hline
\end{tabular}
\end{table}

\paragraph{\textbf{Summary:}} Substituting the word generation layer with a classification layer in Flan-T5 models, although it has no benefits for the SATD identification task, does increase their performance for the SATD classification task, especially when utilizing the smaller versions of the Flan-T5 models. This improvement is likely due to the limited available training data for the SATD classification task.

%%%%%%%%%%%%%%%%%%%%%%%%%%%%%%%%%%%%%%%%%%%%%%%%%%%%%%%%%%%%%%%%%%%%%%

\subsection{\rqfour}

\subsubsection{Approach}
\paragraph{\textbf{SATD Classification.}} In RQ1 to RQ3, we evaluated large language models on SATD identification and classification tasks, using code comments as input. While the Maldonado-62k dataset lacks additional features, the OBrien dataset includes informative features that can enhance model accuracy in predicting SATD categories. Undoubtedly, the code comment itself is the most significant feature in determining the category of an SATD. However, contextual features like file path or surrounding code can aid in a more comprehensive understanding of a code comment. For instance, the presence of the word `test' in the file name or path may increase the likelihood of a code comment being related to a test SATD. Therefore, in this research question, we aim to investigate how effectively LLMs leverage additional contextual features for SATD classification. More specifically, RQ4 compares models using four distinct combinations of input data: 1) just the comment text (with results adopted from RQ1 and RQ2); 2) the file path along with the comment text; 3) the file path, the containing method's signature, and the comment text; 4) the file path, comment text, and the entire containing method, which includes both the signature and body. Given that the containing method can be extensive, it is positioned at the end of the context. This arrangement ensures that in instances of long input data leading to truncation, the critical comment text is preserved. Table~\ref{tab:obrien-sample-input-data} provides an example for each input data combination.

\begin{table}
\caption{A sample for each combination of input data in OBrien dataset}
\label{tab:obrien-sample-input-data} 
\resizebox{12cm}{!}{
\begin{tabular}{|p{0.36\linewidth}|p{0.64\linewidth}|}
\hline
\textbf{Input data} & \textbf{Example} \\
\hline
Comment text & \texttt{TODO: Future work: how to handle gracefully minus numbers} \\
\hline
File path + & file path: \texttt{test/torch\_test.py} \\
Comment text & Technical debt comment: \texttt{TODO: Future work: how to handle gracefully minus numbers} \\
\hline
File path + & file path: \texttt{test/torch\_test.py} \\
Containing method’s signature + & Containing method signature: `````` \texttt{test\_mpc\_sum(self)} """ \\
Comment text & Technical debt comment: \texttt{TODO: Future work: how to handle gracefully minus numbers} \\
\hline
File path + & file path: test/torch\_test.py \\
Comment text + & Technical debt comment: `````` \texttt{TODO: Future work: how to handle gracefully minus numbers} """ \\
Containing method & Containing method body: `````` \\
 & \texttt{def test\_mpc\_sum(self):} \\
 & \texttt{  self.mpc\_sum(3, 5)} \\
 & \texttt{  self.mpc\_sum(4, 0)} \\
 & \texttt{  self.mpc\_sum(5, -5)} \\
 & \texttt{  \# self.mpc\_sum(3, -5) TODO: Future work: how to handle gracefully minus numbers} \\
 & \texttt{  self.mpc\_sum(2 ** 24, 2 ** 12)} """ \\
\hline
\end{tabular}
}
\end{table}

\subsubsection{Results}
\paragraph{\textbf{SATD Classification.}} Table~\ref{tab:obrien-accuracy-per-input-data} presents the results for the SATD classification task using three different approaches: ICL with Flan-T5-XXL\footnote{To create the prompt for the ICL approach, we followed the same method as presented in RQ2 approach. We selected the best result, which was obtained when we included the category descriptions and the five most relevant examples using the SentenceTransformer.}, training the CNN and fine-running the selected six LLMs, and an ensemble approach by the fine-tuned Flan-T5-XL. For the fine-tune approach, when the input data includes the file path or the file path along with the method's signature, the two largest models, Flan-T5-large and Flan-T5-XL, significantly improve their performance compared to when we only provide the comment text. In contrast, the ICL approach and the four fine-tuned smaller LLMs and the CNN model show minimal improvement or even deteriorate. Notably, the last column demonstrates how adding both the file path and the entire containing method to the input data complicates the task for all models except Flan-T5-XL. While this input worsens the performance of all models, Flan-T5-XL utilizes it to achieve better results than when using only the comment text. This finding highlights the superior capability of fine-tuning larger models in processing and leveraging complex data for the SATD classification task. The consistent high performance of the Flan-T5-XL model (with scores ranging from 0.62 to 0.648 across different input combinations) led us to explore an ensemble approach. This method assigns the most frequent label across 12 predictions (4 different input data types × 3 runs = 12). The ensemble approach attained an accuracy of 0.668, the highest among all experiments.

\begin{table} %[h]
\caption{Average accuracy obtained by different data as input (Dataset: OBrien, Approach: 10-fold cross validation, number of runs: 3) CT: Comment Text, FP: File Path, CMS: Containing Method's Signature, CM: Containing Method}
\label{tab:obrien-accuracy-per-input-data}
\resizebox{12cm}{!}{
\begin{tabular}{llllll}
\hline\noalign{\smallskip}
         &       &    & \multicolumn{2}{c}{\textbf{Input data}} & \\
\textbf{Approach} & \textbf{Model} & CT & FP+CT  & FP+CMS+CT & FP+CT+CM \\ 
\noalign{\smallskip}\hline\noalign{\smallskip}
ICL & Flan-T5-XXL (11.1B) & 0.572 & 0.572 & 0.529 & 0.459 \\ 
\noalign{\smallskip}\hline\noalign{\smallskip}
Training & CNN & 0.602 & 0.580 & 0.572 & 0.533 \\ 
\noalign{\smallskip}\hline\noalign{\smallskip}
Fine-    & BERT-base-uncased (110M) & 0.576 & 0.578 & 0.567 & 0.496 \\ 
Tuning   & CodeBERT-base (125M) & 0.611 & 0.606 & 0.571 & 0.492 \\ 
& Flan-T5-small (77M) & 0.537 & 0.520 & 0.506 & 0.436 \\ 
& Flan-T5-base (248M) & 0.565 & 0.560 & 0.546 & 0.468 \\ 
& Flan-T5-large (783M) & 0.575 & 0.599 & 0.590 & 0.545 \\ 
& Flan-T5-XL (2.85B) & \textbf{0.620} & \textbf{0.648} & \textbf{0.628} & \textbf{0.635} \\ 
\noalign{\smallskip}\hline\noalign{\smallskip}
Ensemble & fine-tuned Flan-T5-XL & \multicolumn{4}{c}{\textbf{0.668}} \\
\noalign{\smallskip}\hline
\end{tabular}
}
\end{table}

\paragraph{\textbf{Summary:}} Larger models like Flan-T5-large and Flan-T5-XL significantly outperform other smaller models when contextual information are included in the input data for the SATD classification task, with Flan-T5-XL showing superior capability by utilizing complex data effectively. An ensemble approach using the Flan-T5-XL model yielded the highest accuracy, demonstrating the benefits of larger models and varied inputs.

\section{Discussion}\label{sec:discussion}
\subsection{Impact of epoch number}

To train the selected large language models, we set the number of training epochs to eight. After training, we used the model from the final epoch to test its performance, because there was no validation set available to select the best-performing model. In this subsection, we present the performance of the trained model on the test set across epochs for RQ1.

\begin{figure*}
\includegraphics[width=1.0\textwidth]{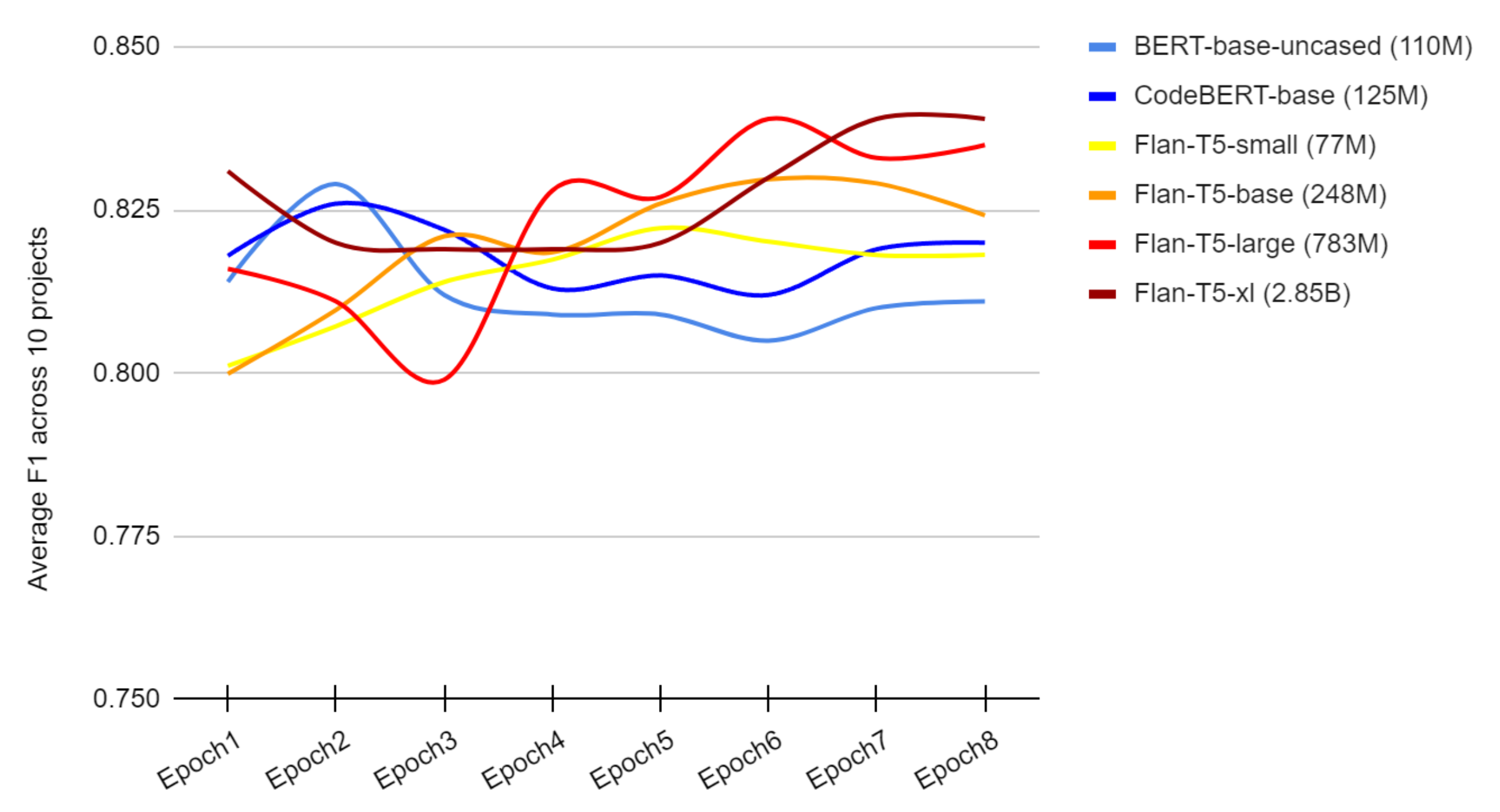}
\caption{Average F1 score across 10 projects in the Maldonado-62k dataset over epochs}
\label{fig:maldonado62k-f1-over-epochs}
\end{figure*}

Figure~\ref{fig:maldonado62k-f1-over-epochs} illustrates the trend of the average F1 score across Maldonado-62k projects from epochs 1 to 8. The figure reveals that all six large language models start with a high performance (approximately 0.81 F1 score) in the first epoch, with subsequent epochs showing only minor fluctuations around this score. This consistency is likely due to the binary classification nature of the SATD identification task and the substantial volume of training data in the Maldonado-62k dataset, allowing these pre-trained models to reach near-optimal performance from the first epoch.

\begin{figure*}
\includegraphics[width=1.0\textwidth]{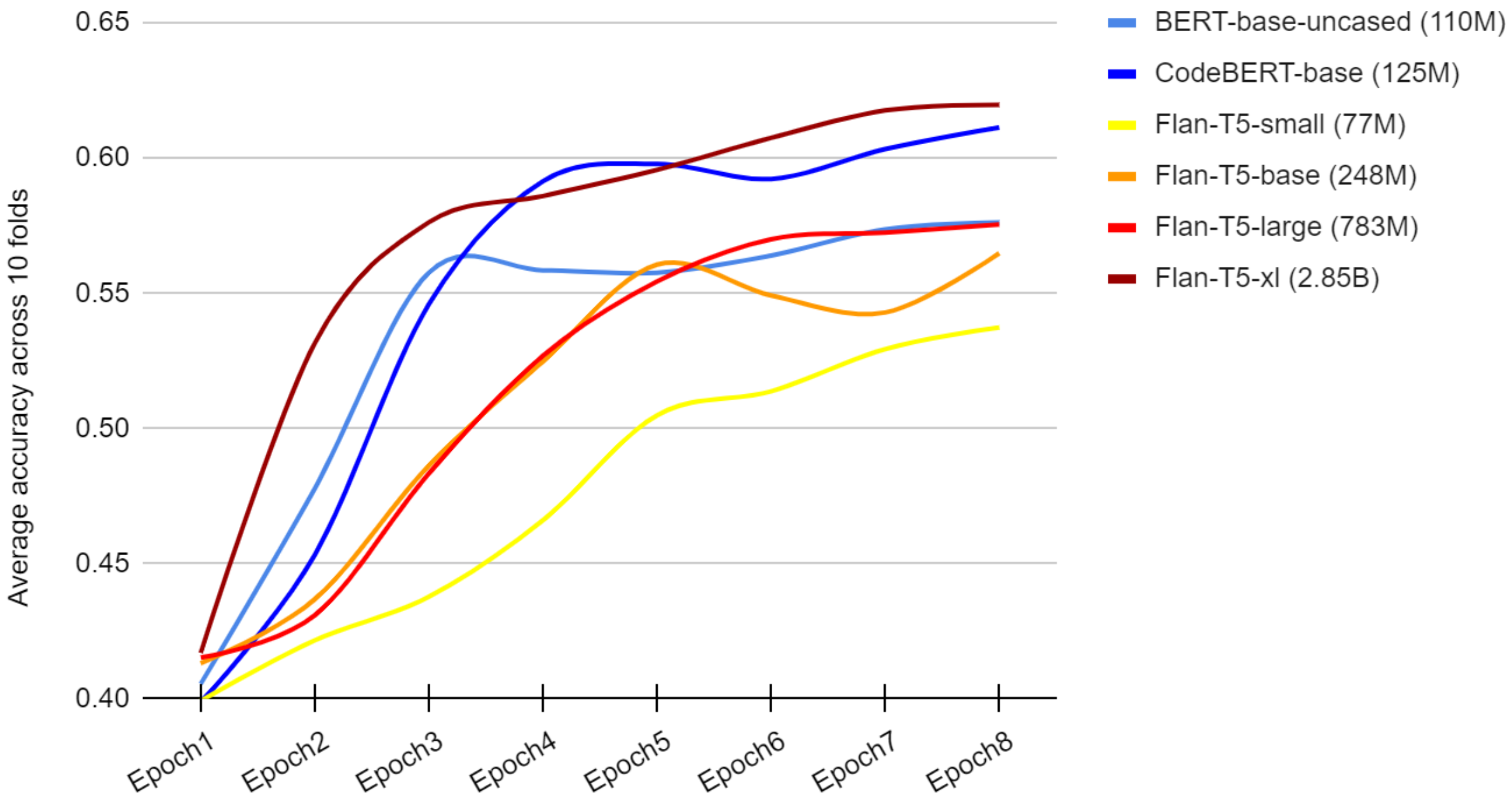}
\caption{Average accuracy across 10 folds over epochs (Dataset: OBrien, Approach: 10-fold cross validation, number of runs: 3)}
\label{fig:obrien-f1-over-epochs}
\end{figure*}

Figure~\ref{fig:obrien-f1-over-epochs} presents the results for the SATD classification task using the OBrien dataset. In contrast to Figure~\ref{fig:maldonado62k-f1-over-epochs}, where smaller models performed competitively with larger models, a more pronounced difference is observable in this figure. Additionally, the performance of all LLMs sharply increases from epoch 1 to epoch 5. Beyond epoch 5, the performance either stabilizes with minor fluctuations or gently continues to increase, suggesting that epoch 8 is an optimal point to stop training.

\section{Threats to Validity}\label{sec:threats}
\textbf{External validity.}
Creating a large dataset with completely error-free labels by human effort is impractical, and the Maldonado dataset is no exception. Although some incorrect labels are not entirely wrong but rather due to different interpretations of the SATD definition, there are numerous instances in this dataset where annotators would choose the opposite label with high agreement. We attempted to mitigate this problem by applying the label modifications proposed in \cite{Yu-2022}, as we found their suggestions reasonable. However, all these proposed modifications address false negative cases. There are also false positives in the dataset, a few of which are reported in \cite{Cassee-2022}. Therefore, we could expect better performance if we had access to a more accurate dataset.

\noindent \textbf{Internal validity.}
In our experiments, we used only two datasets for SATD identification/classification. There are additional datasets for this purpose, but as previously mentioned, they are either created from other software sources such as issue tracking systems \cite{Li-2022} and build systems \cite{Xiao-2021}, which are beyond the scope of this study, or they are derivatives of the Maldonado dataset \cite{Fucci-2021,Cassee-2022}, which we have already utilized. However, due to the lack of available contextual features in this dataset, we were unable to analyze the impact of additional features on this dataset.
Another concern regarding internal validity is the challenge of finding optimal parameters for fine-tuning large language models. We employed a manual investigation approach; however, considering the substantial computational resources required by larger LLMs, thoroughly exploring a wide range of parameter settings is time-consuming. Therefore, we primarily focused on identifying an effective learning rate through manual trials.

\noindent \textbf{Construct validity.}
Regarding SATD identification with the ICL approach (RQ2), we did not find an effective method that improves performance by incorporating examples in prompts compared to simply including SATD-related keywords. There may be approaches that can efficiently select relevant examples from the training set for the specific task of SATD identification.
Another potential threat to the construct validity of our study is the selection of BERT, CodeBERT, and Flan-T5 models for our experiments. There are many more open-source large language models that could have been chosen for this study. The rationale behind selecting Flan-T5 was to investigate the impact of model size on our tasks. Additionally, we faced infrastructure limitations in fine-tuning other models such as LLaMA. Although LLaMA is available in various sizes, its parameter range is significantly larger than that of Flan-T5, spanning from 7 billion to 70 billion parameters.

\section{Conclusion and Future Work}\label{sec:conclusion}
In this study we investigated the effectiveness of large language models for SATD identification and classification. Our results showed that the fine-tuned selected LLMs outperform all SATD identification baselines. For the SATD classification task, while the largest fine-tuned LLM, Flan-T5-XL, still led in performance, the CNN model exhibited competitive performance, even surpassing some LLMs. In both tasks, larger LLMs outperformed smaller counterparts, particularly in SATD classification task that limited training data is available. We also applied the ICL approach on the largest Flan-T5 model, Flan-T5-XXL. For SATD identification, by including different groups of SATD related keywords in the prompts, it provided competitive results with traditional approaches, but couldn't surpass even the smallest fine-tuned LLM. In SATD classification task, incorporating examples and category descriptions in prompts outperforms the zero-shot approach and even surpasses the fine-tuned smaller Flan-T5 models.

Our experiments also showed that substituting the word generation layer with a classification layer in Flan-T5 models increases their performance for the SATD classification task likely due to the limited available training data, especially when utilizing the smaller versions of the Flan-T5 models. 
Finally, we discovered that large fine-tuned models, particularly Flan-T5-XL, effectively utilize additional contextual features, such as surrounding code, to enhance performance. In contrast, smaller fine-tuned models exhibit a decrease in performance when provided with complex contextual information.

One of the notable findings from this study is that the fine-tuned Flan-T5-XL model, despite being trained with human-generated labels from the Maldonado-62k dataset, managed to outperform human annotators in SATD identification, thanks to its pre-trained knowledge. We believe that the primary obstacle to achieving higher performance in large LLMs lies in the quality of the labeled data. Therefore, future research in this domain should focus on preparing data of higher quality, which includes essential details such as commit hashes and file paths. These details provides the location of code comments within repositories and facilitate access to additional information, thereby enhancing the performance of future models.

\begin{acknowledgements}
We acknowledge the support of the Natural Sciences and Engineering Research Council of Canada (NSERC), [funding reference number: RGPIN-2019-05071].
\end{acknowledgements}

\section*{Conflict of Interest}

The authors declare that they have no conflict of interest.

\section*{Data Availability Statements}

The results, source code, and data related to this study are available at \url{https://github.com/RISElabQueens/SATD_LLM}

\bibliographystyle{apalike}      % sadegh: as the basic style doesn't work properly
\bibliography{main}   % name your BibTeX data base

% % Non-BibTeX users please use
% \begin{thebibliography}{}
% %
% % and use \bibitem to create references. Consult the Instructions
% % for authors for reference list style.
% % Format for Journal Reference
% Author, Article title, Journal, Volume, page numbers (year)
% Author, Book title, page numbers. Publisher, place (year)
% \end{thebibliography}

\end{document}